\documentstyle{l-aa}

\input epsf 
\newcommand{\bez}{\begin{eqnarray*}}
\newcommand{\eez}{\end{eqnarray*}}
\newcommand{\be}{\begin{equation}}
\newcommand{\ee}{\end{equation}}

\newcommand{\beq}{\begin{eqnarray}}
\newcommand{\eeq}{\end{eqnarray}}
\newcommand{\bc}{\begin{center}}
\newcommand{\ec}{\end{center}}

\def\cminvcube{\,{\rm cm}^{-3}}
\def\secinv{\,{\rm s}^{-1}}

\def\CR{C_{\rm R}}
\def\ee{$e^+ e^-$\ }

\def\g{$\gamma$}
\def\gcm{\gamma_{\rm cm}}

\def\me{m_{\rm e}}

\def\sigmat{\sigma_{\rm T}}
\def\sigmaann{\sigma_{\rm ann}}
\def\sigmagg{\sigma_{\gamma \gamma}}

\def\xcm{x_{\rm cm}}

 \mathcode`*="002A

\newbox\grsign \setbox\grsign=\hbox{$>$} \newdimen\grdimen \grdimen=\ht\grsign
\newbox\simlessbox \newbox\simgreatbox \newbox\simpropbox
\setbox\simgreatbox=\hbox{\raise.5ex\hbox{$>$}\llap
     {\lower.5ex\hbox{$\sim$}}}\ht1=\grdimen\dp1=0pt
\setbox\simlessbox=\hbox{\raise.5ex\hbox{$<$}\llap
     {\lower.5ex\hbox{$\sim$}}}\ht2=\grdimen\dp2=0pt
\setbox\simpropbox=\hbox{\raise.5ex\hbox{$\propto$}\llap
     {\lower.5ex\hbox{$\sim$}}}\ht2=\grdimen\dp2=0pt

\begin{document} 
 
   \thesaurus{12         
              (02.12.3;  
               02.18.6;  
               13.07.3)} 

   \title{A simple formula for the thermal pair annihilation line emissivity}
  

   \author{Roland Svensson, Stefan Larsson and Juri Poutanen
          }

   \offprints{Roland Svensson}

   \institute{Stockholm Observatory, S--133 36 Saltsj\"obaden, Sweden \\
      Internet: svensson@astro.su.se, stefan@astro.su.se, juri@astro.su.se 
             }

   \date{}

   \maketitle 

\markboth{Svensson, R. et al.: A simple formula for the thermal  
pair annihilation emissivity}{}

\begin{abstract}

We introduce a simple and convenient fitting formula 
for the thermal annihilation line from pair plasmas in cosmic sources. 
The fitting formula is accurate to 0.04\%
and is valid at all photon energies and temperatures
of interest. The commonly used Gaussian line profile
is not a good approximation for broader lines.

 
\end{abstract}
 
\section{Introduction}
\label{sec:intro}

For two reasons a 
need has developed to be able to compute the annihilation line
emissivity in a thermal plasma both exactly and rapidly. 
First, relatively broad and resolved
annihilation lines have been observed in the galactic black hole 
candidates Nova Muscae (Gilfanov et al. 1993), 
and 1E 1740.7-2942 (Bouchet et al. 1991) as well as from
a source observed with {\it HEAO-1} A-4 (Briggs 1991).
With some exceptions (e.g. Macio\l ek-Nied\'zwiecki \& Zdziarski 1994)
a Gaussian line profile is normally used
when fitting the observations. A Gaussian profile does not have the correct 
shape when the temperature becomes sufficiently large. 
Future gamma ray missions such as 
{\it INTEGRAL} will be able to determine the annihilation line shape to such 
accuracy that the exact annihilation line shape must be used 
when fitting the line.
Second, the theoretical modelling of radiative transfer in
the hot thermal plasmas thought to be responsible for the X-ray and \g-ray
emission from e.g. active galactic nuclei, galactic black hole candidates,
and gamma-ray bursts requires a fast way to evaluate the thermal
annihilation line emissivity.

The thermal annihilation line emissivity 
is exactly determined by a single, one-parameter integral 
first derived by Svensson (1983, see also Dermer 1984). 
Svensson (1983) also gave some less convenient but 
very accurate polynomial fits to that integral. 
Our aim here is to provide a single, simple, but accurate 
fitting formula valid at all 
temperatures and photon energies of interest. 
This fitting formula can then be used when modelling 
observations. It also allows rapid evaluations of 
the annihilation line emissivity in 
radiative transfer calculations.

The structure of the paper is as follows. In \S~\ref{sec:theory},
we summarize previous theoretical results regarding
the thermal annihilation line emissivity
and in \S~\ref{sec:formula} we introduce a very accurate fitting formula
for the line shape. 
Finally, in \S~\ref{sec:discussion}, we summarize our results. 

\section{Theory of the thermal pair annihilation line}
\label{sec:theory}

Consider a plasma consisting of electrons and positrons of densities
$n_-$ and $n_+$, respectively. The electrons and positrons are assumed to 
have Maxwell-Boltzmann energy distributions at the same temperature, $T$.
Svensson (1983) showed using simple detailed balance
arguments that the emission rate, 
$(dn/dt)(x) dx \cminvcube \secinv$, of annihilation 
photons of dimensionless energy, 
$x \equiv h \nu / \me c^2$, in the energy interval $dx$
in such a plasma is given 
by an integral over the cross section for the reverse process,
photon-photon pair production:
\beq
\lefteqn{\frac{dn}{dt}(x,\theta)dx =
n_+n_- c dx \times} \nonumber\\
&&\hspace{1cm}
\frac{2}{\theta \left[ K_2(1/\theta) \right]^2}
e^{-x/\theta}
\int_1^{\infty} ds s \sigmagg (s) e^{-s/x\theta}\,.
\label{eq:svensson}
\eeq
Here, $K_2(1/\theta)$ is the modified Bessel function of second kind of
order 2, the dimensionless temperature $\theta$ is defined as
$\theta \equiv kT / \me c^2$, and $\sigmagg (s)$
is the cross section for photon-photon pair production, 
$\gamma \gamma \rightarrow e^+e^-$ (Jauch \& Rohrlich 1976),
\beq
\lefteqn{\sigmagg (s)
= \sigmat \frac{3}{8s}
\left[ \left( 2+ \frac{2}{s} - \frac{1}{s^2}\right)
\ln \left( \sqrt{s} + \sqrt{s-1}\right) \right.} \nonumber \\
&&\hspace{1cm}
- \left. 
\left( 1+ \frac{1}{s} \right) \left( 1 - \frac{1}{s}\right)^{1/2}
\right]\,,
\label{eq:ggcross}
\eeq
where $s \equiv \xcm^2$ with $\xcm$ being the photon energy in the 
center-of-momentum frame. Near pair production threshold, $s-1 \ll 1$,
we have $\sigmagg (s) = \sigmat (3/8)(s-1)^{1/2}$, while far 
above the threshold, $s \gg 1$, we have 
$\sigmagg (s) = \sigmat (3/8s)(\ln 4s - 1)$.
 
Dermer (1984) developed a general theory for production spectra from
binary collisions in thermal
plasmas by reducing a six-dimensional integral over the differential cross
section to a double integral. Dermer applied this result to 
the pair annihilation process,
in which case the double integral reduces to a single integral over
the pair annihilation cross section:
\begin{eqnarray}
\lefteqn{\frac{dn}{dt}(x,\theta)dx =
n_+n_- c dx \frac{2}{\theta \left[ K_2(1/\theta) \right]^2}
\times} \nonumber \\
&&\hspace{1cm} e^{-x/\theta} \int_1^{\infty} ds 2(s-1)
\sigmaann (s) e^{-s/x\theta}\,.
\label{eq:dermer}
\end{eqnarray}
Here, $\sigmaann (s)$ is the pair annihilation cross section,
and $s \equiv \gcm^2$, with $\gcm$ being the particle Lorentz
factors in the center-of-momentum frame. We have rewritten Dermer's
integral in terms of $s$ instead of his chosen integration variable, 
$\gamma_{\rm r}= 2s+1$, i.e. the Lorentz factor of one of the
particles in the rest frame of the other.
The cross sections for the annihilation process, $\sigmaann$, 
and its inverse, $\sigmagg$ are related to each other
due to microscopic detailed balance:   
\begin{equation}
\lefteqn{\sigmaann (s) = \frac{s}{2(s-1)} \sigmagg (s)\,;
\quad \qquad \xcm = \gcm\,.}
\label{eq:anncross}
\end{equation}
Using this relation we find that the two expressions 
(\ref{eq:svensson}) and (\ref{eq:dermer})  
for the annihilation emissivity, 
obtained using two quite different methods, are identical.

We now need too evaluate the Bessel function factor as a function of
$\theta$ and the integral as a function of the parameter $x \theta$.
To an accuracy of 0.06\% the Bessel function factor can be approximated
with (Svensson 1983)
\vspace{3mm}
\begin{eqnarray}
\lefteqn{
\left[ K_2(1/\theta) \right]^2 =
4 \theta^4 e^{-2/\theta} \times} \nonumber \\
&&\hspace{1cm} \left[ 1+ 2.0049 \theta^{-1}
+ 1.4774 \theta^{-2} + \pi (2 \theta)^{-3} \right]\,,
\label{eq:bessel}
\end{eqnarray}
which has the correct analytic behavior for $\theta \ll 1$ and $\theta \gg 1$.
Using the two asymptotic forms for the cross section at $s \ll 1$
and $s \gg 1$ allows the integral to be solved analytically in the two limits 
$x\theta \ll 1$ and $x\theta \gg 1$ giving
\vspace{3mm}
\begin{eqnarray}
\lefteqn{
\int_1^{\infty} ds s \sigma_{\gamma \gamma} (s) e^{-s/x\theta}
= } \nonumber\\
&&\hspace{1cm}
\left\{
\begin{array}{ll}
\sigmat \frac{3}{16} \pi^{1/2} (x\theta)^{3/2}
e^{-1/x\theta}\,;\quad & x\theta \ll 1\,, \vspace{2mm} \\
\sigmat \frac{3}{8} x\theta (\ln 4 \eta x\theta - 1)\,;
& x\theta \gg 1\,, 
\end{array}
\right.
\label{eq:integral}
\end{eqnarray}
where $\eta = \exp( - \gamma_{\rm E})$ = 0.56146.., $\gamma_{\rm E}$ 
being Euler's constant.
Using the asymptotic analytic forms of both the Bessel function 
factor and the 
integral gives the two following useful asymptotic expressions for
the annihilation emissivity
\begin{eqnarray}
\lefteqn{\frac{dn}{dt}(x,\theta)dx =
n_+n_- c \sigmat dx \times} \nonumber \\ 
&&\hspace{0.5cm}
\left\{
\begin{array}{ll}
\frac{3}{4 ( \pi \theta)^{1/2}}
x^{3/2} \exp\left[\frac{-(x-1)^2}{x\theta} \right]\,;  
& x\ll \frac{1}{\theta}\,,\,\,\, \theta \ll 1\,,  \vspace{2mm}\\
\frac{3}{16\theta^4}(\ln 4 \eta x\theta -1) x e^{-x/\theta}\,;
& x\gg \frac{1}{\theta}\,,\,\,\, \theta \gg 1\,. 
\end{array} 
\right.
\label{eq:asymptotics}
\end{eqnarray}
At relativistic temperatures, $\theta \gg 1$, 
the kinetic energy of the annihilating
pair dominates over the rest mass energy, and the annihilation
spectrum mimics a Maxwellian with a peak at $x \approx \theta$.
At nonrelativistic temperatures, $\theta \ll 1$, the rest mass energy 
dominates and the annihilation line peaks at $x \approx 1 + 3 \theta /4$
with the slight blue shift and broadening being due to 
the kinetic energy of the annihilating particles (Ramaty \& M\'esz\'aros
1981). The nonrelativistic expression is valid through out the 
whole low energy wing of the line, and far into the high energy wing
up to a photon energy of $1/\theta$. If one only considers the line core
near $x \sim 1$, or, more precisely $x-1 \ll 1$, one gets
\begin{eqnarray}
\lefteqn{\left( \frac{dn}{dt}\right)_{\rm G}(x,\theta)dx =
n_+n_- c \sigmat dx \frac{3}{4 ( \pi \theta)^{1/2}}
e^{-(x-1)^2/\theta};} \nonumber \\
&&\hspace{4cm} 
x-1 \ll 1, \quad \theta \ll 1\,,
\label{eq:gaussian}
\end{eqnarray}
which shows that the line core has a Gaussian shape
for $\theta \ll 1$. It is 
this nonrelativistic line core approximation  
that is normally used when fitting observed annihilation features.
Then, however, an exponent of $(x-x_{\rm c})^2/\theta$ is used with 
{\it two} fitting parameters, the line center energy, $x_{\rm c}$, and the
temperature, $\theta$. 

It is appropriate at this point to briefly mention other 
work on the annihilation line profile. Aharonian, Atoyan, \& 
Sunyaev (1983) also derived an single
integral expression for the thermal annihilation line emissivity,
but this differs from the self consistent and correct expressions 
(\ref{eq:svensson}) and ({\ref{eq:dermer})
obtained by Svensson (1983) and Dermer (1984). While analytical 
(or numerical) integration of the correct expression over photon energy 
gives the well known analytical expression for the annihilation rate
already obtained by Weaver (1976), a similar integration over
Aharonian et al.'s expression does not.
The reason for the discrepancy 
is the second term $I_2$ in the integrand of their
equation (A.4) originating from the term $\propto \cos^2 \chi'$ in their
equation (33). Using $I_1$ only, gives the correct analytical result.
   
The multi-dimensional integral for the emissivity that Dermer started out
with in his derivation has been evaluated numerically using
Monte Carlo methods by Zdziarski (1980), Ramaty \& M\'esz\'aros (1981),
and Yahel (1982). The results of the first two papers agree
with the correct analytical results, while the calculations by Yahel
give quite a different line shape. The reason for the 
discrepancy is due to Yahel confusing the angle between the annihilation 
photon and the particle momentum 
with the angle between the direction of transformation to the
lab frame and the particle momentum, all in the center-of-momentum frame. 
The existence of simple analytical expressions 
for the thermal annihilation emissivity has now eliminated the need 
for the time-consuming Monte Carlo calculations appearing in these
papers.

\section{An excellent fitting formula}
\label{sec:formula}

We now attempt to provide fitting formulas for the integral in 
equation (\ref{eq:svensson}) valid for all values of $x\theta$.
As the annihilation line is swamped by bremsstrahlung emission for
$\theta$ larger than about 3 (Svensson 1982, Macio\l ek-Nied\'zwiecki 
et al. 1995), a useful formula for fitting observable
lines would need to be valid for, say, $x\theta < 20$.
We therefore write the integral as 
\begin{eqnarray}
\lefteqn{\int_1^{\infty} ds s \sigmagg (s) e^{-s/x\theta}
= \sigmat \frac{3}{16} \pi^{1/2} (x\theta)^{3/2}
e^{-1/x\theta} C(x \theta)\,;} \nonumber \\
&&\hspace{5cm} x\theta < 20\,,
\label{eq:corrdef}
\end{eqnarray}
where $C(y)$ is the correction factor to the asymptotic expression
(\ref{eq:integral}) for $x\theta \ll 1$. 
The annihilation emissivity, equation 
(\ref{eq:svensson}) then becomes
\vspace{3mm}
\begin{eqnarray}
\lefteqn{\frac{dn}{dt}(x,\theta)dx =
n_+n_- c \sigmat dx
\frac{3}{8} (\pi \theta)^{1/2} x^{3/2} \times
}\nonumber \\
&&\exp \left(- \frac{x+x^{-1}}{\theta} \right)
\frac{C(x\theta)}{\left[ K_2(1/\theta) \right]^2}\,;
\quad {\rm any} \; \theta, \,\,x\theta < 20\,.
\label{eq:fitform}
\end{eqnarray}
The Bessel function factor is given by the approximative equation
(\ref{eq:bessel}) while $C(y)$ was computed to an accuracy of $10^{-5}$
using integration routines from Numerical Recipes (Press et al 1986).
The resulting $C(y)$ is shown in the top panel in Figure~\ref{fig:corrfactor}.
In order to find accurate fitting formulas for $C(y)$,
we use two different Pad\'e approximations with increasing
accuracy.
Fitting the numerical results of Figure~{\ref{fig:corrfactor}
with the Pad\'e approximation
\begin{equation}
C(y)= \frac{1 + a_1 y + a_2 y^2}{1+ b_1 y + b_2 y^2}\,;
\qquad  y < 20\,,
\label{eq:padefit1}
\end{equation}
gave the following coefficients: $a_1= 7.308$, $a_2 = 0.4946$, 
$b_1=5.002$, and
$b_2=0.8075$. The error as a function of $y \equiv x\theta$ 
is shown in the middle panel
in Figure~\ref{fig:corrfactor}. The maximum error is about 1.4\%
for $x\theta < 20$ and 0.5\% for $x\theta < 10$.
Similarly, using the expression
\begin{equation}
C(y)= \frac{1 + c_1 y + c_2 y^2+ c_3 y^3}{1+ d_1 y + d_2 y^2+ d_3 y^3}\,;
\qquad y < 20\,,
\label{eq:padefit2}
\end{equation}
we find the coefficients $c_1=6.8515487$, $c_2= 1.4251694$,
$c_3=0.017790149$, $d_1=4.63115589$, $d_2=1.5253007$, and
$d_3=0.04522338$.
The error of this fit is shown in the lower
panel of Figure~\ref{fig:corrfactor}, where the maximum error is seen to
be 0.04 per cent. 

\begin{figure}[tb]
\bc
\leavevmode
\epsfxsize=8.8cm 
\epsfbox{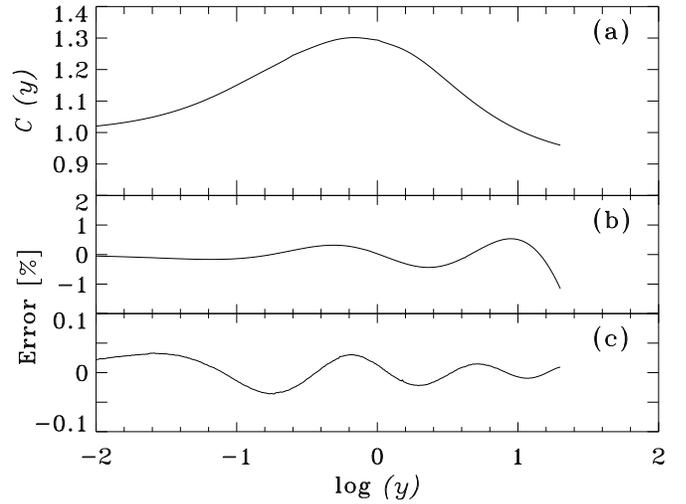}
\ec
 \caption{
(a) The correction factor, $C(y)$, as function of $y \equiv x\theta$ for
$y < 20$. (b) Relative error in per cent using our first Pad\'e
approximation, equation~(11), for $C(y)$.
(c) Relative error in per cent using our second Pad\'e
approximation, equation~(12), for $C(y)$.
}
\label{fig:corrfactor}
\end{figure}

In rare circumstances in numerical simulations one may need to
know the emissivity for $x\theta > 20 $. Applying a correction
factor, $\CR(y)$, to the asymptotic limit of the integral,
equation~(\ref{eq:integral}) for $x\theta \gg 1$, 
we have
\begin{eqnarray}
\lefteqn{\int_1^{\infty} ds s \sigma_{\gamma \gamma} (s) e^{-s/x\theta}
=  \sigmat \frac{3}{8} x\theta (\ln 4 \eta x\theta - 1)
\CR(x\theta)\,;} \nonumber \\
&&\hspace{5cm} 
x\theta > 20\,. 
\label{eq:corrdef2}
\end{eqnarray}
Then the annihilation emissivity becomes
\begin{eqnarray}
\lefteqn{\frac{dn}{dt}(x,\theta)dx =
n_+n_- c \sigmat dx 
\frac{3}{4} x (\ln 4 \eta x\theta -1) \times} \nonumber \\
&&\hspace{1cm} e^{-x/\theta}
\frac{\CR(x\theta)}{\left[ K_2(1/\theta) \right]^2}\,,
\qquad {\rm any} \; \theta\,, \; x\theta > 20\,.
\label{eq:fitform2}
\end{eqnarray}
Fitting the results from numerically evaluating the integral gives
\vspace{3mm}
\begin{eqnarray}
\lefteqn{
\CR(y)= 1+ 2.712 y^{-1} - 55.60 y^{-2} + 1039.8 y^{-3}
}\nonumber \\ 
&&\hspace{1cm}  - 7800 y^{-4}\,; \qquad \qquad 
\qquad {\rm any}\;\theta\,, \; y > 20\,
\label{eq:relfit}
\end{eqnarray}
to an accuracy of 0.1 per cent.

Our finale recipe for computing the annihilation emissivity for
$x\theta < 20 $ is then 
to use equation (\ref{eq:fitform}) together with the fit for $K_2(\theta)$,
equation (\ref{eq:bessel}), and one
of the fits for $C(y)$, equations (\ref{eq:padefit1}) or (\ref{eq:padefit2}).
The total error is at most 0.1 per cent for the latter case.
For those rare cases when one is interested in the annihilation emissivity
for $x\theta > 20 $, one should use
equation (\ref{eq:fitform2}) together with the fits, equations 
(\ref{eq:bessel}) and (\ref{eq:relfit}). 
The total error is at most 0.15 per cent. 
In Figure~\ref{fig:lines}, we show the dimensionless
annihilation line shapes,
$dn/dt(x,\theta)/n_+n_-c\sigmat$, using the recipe given above
for different dimensionless
temperatures, $\theta$.
The line core is approximately Gaussian for temperatures less than about 
$10^9$ K,
but becomes strongly asymmetric for larger temperatures.

\begin{figure}[tb]
\bc
\leavevmode
\epsfxsize=8.8cm 
\epsfbox{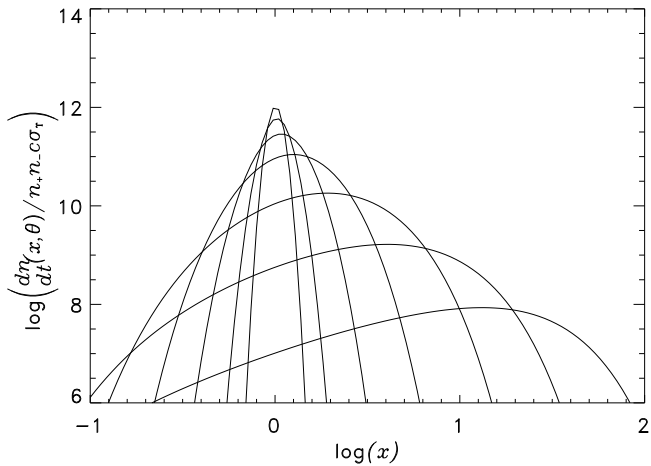}
\ec
\caption{
Dimensionless thermal pair annihilation line shapes,  
$dn/dt(x,\theta)/n_+n_-c\sigmat$, as function of dimensionless
photon energy, $x\equiv h\nu/ \me c^2$,
using the analytical fits in the text for dimensionless
temperatures, $\theta \equiv kT/ \me c^2 =$ 
0.01, 0.03, 0.1, 0.3, 1, 3, and 10 
corresponding to temperatures, $T$, of about
$5 \times 10^7$ K, $1.5 \times 10^8$ K, $5 \times 10^8$ K,
$1.5 \times 10^9$ K, $5 \times 10^9$ K, $1.5 \times 10^{10}$ K,
and $5 \times 10^{10}$ K, or to $kT$ of about
5 keV, 15 keV, 50 keV, 150 keV, 500 keV, 1.5 MeV, and 5 MeV, respectively.
}
\label{fig:lines}
\end{figure}

\section{Summary and Discussion}
\label{sec:discussion}

We have derived very convenient formula to use when fitting
observations of thermal annihilation lines or for the rapid evaluation
of the annihilation line emissivity in radiative transfer
calculations.

The main result is equation (\ref{eq:fitform}) to be used together with  
equations (\ref{eq:bessel}), and (\ref{eq:padefit1}) or (\ref{eq:padefit2})
when computing the annihilation emissivity for
$x\theta < 20 $. Using equation (\ref{eq:padefit2}) gives an error
 of less than 0.1 per cent. The error can be reduced to less than  0.04 per cent
by evaluating $K_2(1/\theta)$ more accurately using, e.g., routines in
Numerical Recipes (Press et al. 1986).
For $x\theta > 20 $, one should use
equations (\ref{eq:fitform2}) together with equations 
(\ref{eq:bessel}) and (\ref{eq:relfit}). The total error is at most 0.15
per cent. 

There are a few limitations to the expressions we give. They assume
a Maxwell-Boltzmann distribution for the annihilating particles.
The mechanism normally thought to be responsible
for maintaining such a thermal distribution
is energy exchange through Coulomb scattering. This fails
for temperatures $kT$ of order $\me c^2$, i.e. $\theta > 1$ as discussed
by e.g. Ghisellini, Haardt \& Fabian (1993).
Other thermalizing mechanisms may, however, be operating. Ghisellini,
Guilbert, \& Svensson (1988) and Ghisellini \& Svensson (1990)
show that synchrotron self-absorption may
act as an extremely efficient thermalizing
mechanism. 

The expressions given are in the Born-approximation. Below temperatures
of about $10^8 $ K, Coulomb corrections must be included.
At such low temperatures, the annihilation takes place through positronium 
formation and one must also include the positronium continuum due to
three-photon annihilation (Ore \& Powell 1949).

\begin{acknowledgements}
This research has been supported by  grants from the 
Swedish Space Board and the Swedish Natural Science Research Council.
J.P. is also grateful for generous support from the 3rd
Compton Symposium.  
\end{acknowledgements}

\end{document}